\documentclass[prl,showpacs,twocolumn,aps]{revtex4}
\usepackage[dvipdfm]{graphicx}
\usepackage[dvipdfm]{color}
\usepackage{dcolumn}
\usepackage{amsmath}
\usepackage{amssymb}

\newcommand{\lco}{$\rm La_2CuO_4$}
\newcommand{\lsco}{$\rm La_{2-x}Sr_xCuO_4$}
\newcommand{\lssco}{$\rm La_{1.983}Sr_{0.017}CuO_4$}

\newcommand{\lczfo}{$\rm La_{2}Cu_{0.85}Zn_{0.15}O_4$}
\newcommand{\lsczo}{$\rm La_{2-x}Sr_xCu_{1-z}Zn_zO_4$}
\newcommand{\lssczo}{$\rm La_{1.983}Sr_{0.017}Cu_{0.9}Zn_{0.1}O_4$}

\newcommand{\HSF}{$H_{SF}$}

\newcommand{\DM}{Dzyaloshinsky--Moriya}

\begin{document}
\title{Frustration of the interlayer coupling by mobile holes in $\bf
La_{2-x}Sr_xCuO_4$ ($\bf x<0.02$)}
\author{M.~H\"ucker$^{1,4}$, H.-H.~Klauss$^{2}$, and B.~B\"uchner$^{3}$}
\affiliation{$^1$Physics Department, Brookhaven National Laboratory, Upton, New York 11973, USA}
\affiliation{$^2$Metallphysik und Nukleare Festk\"orperphysik, TU-Braunschweig, 38106 Braunschweig, Germany}
\affiliation{$^3$II. Physikalisches Institut, RWTH--Aachen, 52056 Aachen, Germany}
\affiliation{$^4$II. Physikalisches Institut, Universit\"at zu K\"oln, 50937 K\"oln, Germany}
\date{\today}
\begin{abstract}
We have studied the interlayer coupling in the antiferromagnetic (AF) phase of Sr and Zn
doped \lco\ by analyzing the spin flip transition in the magnetization curves. We find
that the interlayer coupling strongly depends on the mobility of the hole charge
carriers. Samples with the same hole content as well as the same N\'eel temperature but a
different hole mobility, which we adjusted by Zn co-doping, can have a very different
interlayer coupling. Our results suggest that only mobile holes can cause a strong
frustration of the interlayer coupling.
\end{abstract}
\maketitle
In \lsco\ the 3D AF order is destroyed by a remarkably small amount of 2\% holes
($x=0.02$)~\cite{Niedermayer98}. In this compound the S=1/2 Cu spins in the $\rm CuO_2$
planes form a 2D Heisenberg AF with a strong Cu--Cu in-plane exchange of $J=135$~meV, and
a very weak interlayer coupling $J_\perp \sim 10^{-5}J$. Though the interlayer coupling
is indispensable to establish the 3D AF order~\cite{Aharony88}, almost nothing is known
about its dependence on doping with holes and/or static spin vacancies. For \lco\
Thio~et~al. have demonstrated that $J_\perp$ can be determined from a spin flip (SF) for
$H\parallel c$ due to the \DM\ (DM) Cu--spin canting~\cite{Thio88}.

In recent years, great efforts have been made to map each stage of the suppression of the
3D order which is a precondition for the occurrence of superconductivity
(SC)~\cite{Niedermayer98,Chou93,Borsa95,Huecker98c}. The N\'eel temperature $T_N$
decreases from $325$~K for $x=0$ to about 80~K for $x=0.019$ and then drops to zero
within a Sr range of $\Delta x\sim 0.001$ around $x=0.02$~\cite{Niedermayer98,
Huecker98c}. At the same time for $x\gtrsim 0.008$ the so called spin freezing regime
(SFR) evolves at $T< 30$~K~\cite{Chou93,Borsa95}. At $x=0.02$ the SFR crosses over to the
cluster spin glass phase which reaches into the SC phase that appears at $x=
0.06$~\cite{Niedermayer98}. In the AF phase for $30 {\rm K} < T < T_N$ $\rm \mu$SR and
NQR probe a rapid decrease of the internal field $H_{int}$ with increasing $x$. However,
in the SFR $H_{int}$ again rises and independent of $x$ reaches almost the value in \lco
. Recent neutron diffraction data prove the SFR to have a incommensurate AF spin
modulation, which coexists with the commensurate AF order~\cite{Matsuda01a}.

Various models were suggested to describe the suppression of the AF
order~\cite{Aharony88,Cho93,Korenblit99,Suh98a,Neto98b}. The frustration
model~\cite{Aharony88} assumes that individual localized holes cause a frustration of the
in-plane and the interlayer coupling. In the finite size scaling model~\cite{Cho93} holes
segregate into domain walls which limit the 2D correlation length $\xi_{2D}\propto
\sqrt{T_N}$. The SFR was associated with the breakup of the domain walls when holes
localize at low T~\cite{Borsa95}. In contrast, recent neutron diffraction data support
the existence of magnetic domain walls in particular for the SFR~\cite{Matsuda01a}.

To elucidate crucial parameters for the suppression of the 3D AF order in \lsco , we have
studied the interlayer coupling in Sr and/or Zn doped samples. Zn doping is used to
reduce the mobility of the holes as well as to introduce spin vacancies. From our
analysis of the SF transition we find that only mobile holes cause a drastic suppression
of the interlayer coupling. Co-doping with Zn recovers an interlayer coupling almost as
strong as in pure \lco\ due to the localization of holes. Surprisingly, samples with
similar $T_N$ can have a completely different interlayer coupling, depending on the hole
mobility and the number of spin vacancies.

%\section{Experimental}

The DC magnetization $M(H)$ of five polycrystals \lsczo\ (Tab.~\ref{tab1}) was measured
using a VSM ($T_{max}=290$~K, $H_{max}=14$~Tesla). Samples were annealed in vacuum (1/2h,
800$^o$C), their preparation was described in Ref.~\onlinecite{Huecker98c}.

\begin{figure}[t]
\center{\includegraphics[width=0.95\columnwidth,angle=0,clip]{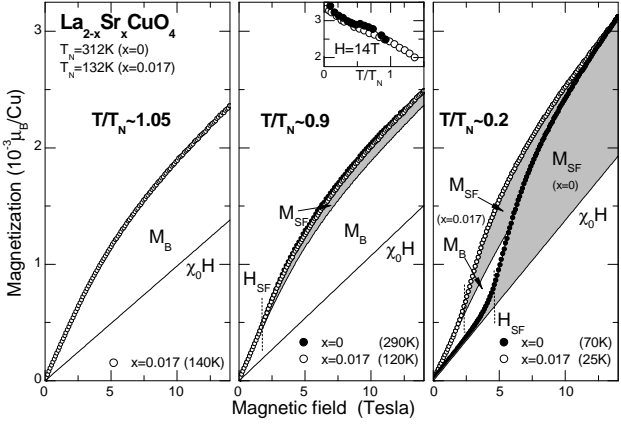}}
\caption[]{Magnetization curves $M(H)$ of \lsco\ with $x=0$ and $x=0.017$ at reduced
temperatures $T/T_N=1.05$, 0.9, and 0.2 (see text). Inset: $M(14 {\rm T})$ vs $T/T_N$.
For $x=0$ data only for $T/T_N<1$ as $T_N>T_{max}(VSM)$.} \label{fig1}
\end{figure}

\footnotesize
\begin{table}[b]
\begin{center}
\begin{tabular}{l|l|c|c|c|c}
Sr    &  Zn
&$T_N$(K)&$H_{SF}$(T)&$M_{DM}^{AF}$($10^{-3}\rm\frac{\mu_B}{Cu}$)&$ J_\perp$($\mu$eV)\\
\hline\hline
0     &  0    &  312  &  4.5(5)  &  2.7(3)                        &  2.9(5)       \\
0.011 &  0    &  222  &  3.6(3)  &  2.1(3)                        &  1.7(5)       \\
0.017 &  0    &  132  &  2.4(2)  &  1.2(2)                        &  0.7(3)       \\
0.017 &  0.10 &  134  &  3.9(4)  &  3.0(3)                        &  2.7(5)       \\
0     &  0.15 &  166  &  3.6(4)  &  3.4(3)                        &  2.7(5)       \\
\end{tabular}
\end{center}
\caption[]{Studied \lsczo\ samples (see text).} \label{tab1}
\end{table}
\normalsize
%
%\section{Results and analysis}
%
%\subsection{Cu spin canting in $\bf La_{2-x}Sr_xCuO_4$ polycrystalls}
%
In the LTO phase of \lco\ the $\rm CuO_6$-octahedra are tilted. As a consequence, nearest
neighbor Cu spins are slightly canted against each other due to a DM superexchange
term~\cite{Thio88}. Below $T_N$ the canted moments $M_{DM} \parallel c$ of adjacent
planes are AF ordered, but can be flipped for $H\parallel c$. Because of this moments of
the order of $\sim 10^{-3}\mu_B/$Cu the $M(H)$ curves are non-linear in the AF phase as
well as in the paramagnetic (PM) phase~\cite{Thio88,Thio94}. This can be seen in
Fig.~\ref{fig1} where we compare $M(H)$ of \lco\ ($T_N=312$~K) and \lssco\ ($T_N=132$~K)
as a function of $T/T_N$. In the PM phase ($T/T_N>1$) we find that the non-linear
contribution can be described by a Brillouin term $M_B$. In the AF phase ($T/T_N<1$) the
non-linear contribution is always a combination of $M_B$ and a term $M_{SF}$ which arises
from the SF of the DM moments:
\begin{equation}\label{equ_m}
M = \left\{ \begin{array}{l@{\quad}l}
\chi_0 H + M_{B}          & \quad\quad (T > T_N)  \\
\chi_0 H + M_{B} + M_{SF} & \quad\quad (T < T_N)\
\end{array} \right.
\end{equation}
where $\chi_0 H$ accounts for all linear terms. As we will see below, $M_{SF}$ is a
function of the long range AF ordered fraction of the DM moment $M_{DM}^{AF}(T) \propto$
AF order parameter, and $M_{B}$ is a function of the non-AF-ordered (NO) fraction
$M_{DM}^{NO}$, both per Cu spin. Just below $T_N$ $M(H)$ for $x=0$ and $x=0.017$ is
almost identical up to 14~Tesla and $M_{SF}$ is very small (grey shaded) while $M_B$ is
still large. In contrast, at low $T$ the non-linear part of $M(H)$ strongly depends on
$x$. For $x=0$ it is dominated by $M_{SF}$ whereas $M_B \simeq 0$. For $x=0.017$ $M_{SF}$
is much smaller and of about the same order as $M_{B}$. Clearly visible is also a much
larger $H_{SF}$ for $x=0$.  At 14~Tesla $M$ is almost $x$-independent for all $T/T_N$
(inset Fig.~1).

%
%

%\subsection{Polycrystal analysis of the spin flip}
%

\begin{figure}[t]
\center{\includegraphics[width=0.94\columnwidth,angle=0,clip]{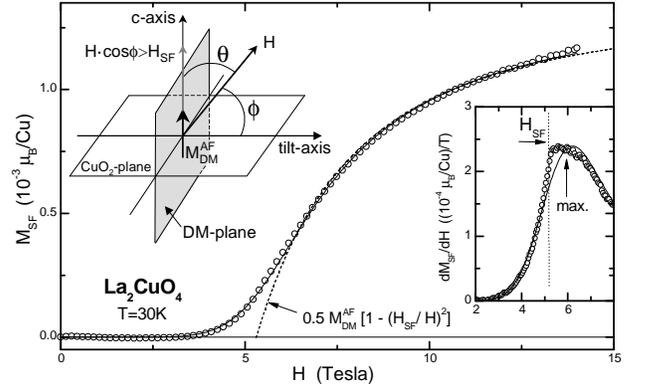}} \caption[]{Spin
flip term $M_{SF} = M-M_B-\chi_0 H$ in \lco\ at 30~K. (-~-~-)/(-----) fit according to
Eq.~\ref{equ_mdm1} without/with gaussian distribution of $H_{SF}$. Left inset: Spin flip
takes place when $H cos \theta \parallel c$ exceeds $H_{SF}$. Only $M_{DM}^{AF} cos
\theta \parallel H$ contributes to $M_{SF}$. Right inset: Derivative $dM_{SF}/dH$.}
\label{fig2}
\end{figure}

The analysis of the SF enables to extract $H_{SF}(T)$ and $M_{DM}^{AF}(T)$. From the low
$T$ limes $J_\perp$ can be calculated~\cite{Thio88}:
\begin{equation}\label{equ_flip}
M_{DM}^{AF}(0) H_{SF}(0) \simeq S^2 J_\perp\ .
\end{equation}
The SF takes place when $H$ acting on $M_{DM}^{AF}$ (see left inset Fig.~\ref{fig2})
overcomes the interlayer coupling, i.e. for $H cos \theta \geq H_{SF}\parallel c$ all
$M_{DM}^{AF}$ become ferromagnetically aligned $\parallel c$. In a single crystal ($H
\parallel c$) the SF causes a step like increase of $M(H)$ at $H\simeq H_{SF}$ by
$M_{SF}=M^{AF}_{DM}$. In a polycrystal the crystallites are oriented randomly. If
$H_{SF}$ is identical in all crystallites we obtain the following field dependence of
$M_{SF}$ for $H \geq H_{SF}$ from integration over all directions:
\begin{equation}\label{equ_mdm1}
M_{SF}(H) = 1/2 M_{DM}^{AF}[1-(H_{SF}/H)^2].
\end{equation}
Obviously in a polycrystal $M_{SF}$ converges to $1/2 M^{AF}_{DM}$ for $H\rightarrow
\infty$. In Fig.~\ref{fig2} we show $M_{SF} = M-M_B-\chi_0 H$ for \lco\ at 30~K. For
large $H$ Eq.~\ref{equ_mdm1} (dashed line) yield a good fit to the data. Assuming a
gaussian distribution of $H_{SF}$ in the polycrystal, integration of Eq.~\ref{equ_mdm1}
over $H_{SF}$ yields the solid line in Fig.~\ref{fig2} which perfectly fits the data. The
extracted parameters $M_{DM}^{AF}= 2.55 \times 10^{-3}$~$\rm \mu_B$ and
$H_{SF}=5.2$~Tesla are in fair agreement with single crystal data in Ref.~\cite{Thio88}.
Note, that in a polycrystal $H_{SF}$ is located slightly below the maximum of $dM/dH$
(inset).

For the Brillouin term $M_B$ in the PM phase of a polycrystal a similar geometrical
consideration as for Eq.~\ref{equ_mdm1} yields the phenomenological formula:
\begin{equation}\label{equ_mdmx}
M_{B}(H) = M_{DM}^{NO} \int^{\pi/2}_{0} tanh(k H sin \phi)\ sin^2 \phi\ d\phi.
\end{equation}
where $k=M_{DM}^{NO} N/(k_B T + J_\perp N^2 S^2)$ is a phenomenological expression with
$N=(\xi_{2D}/a)^2$ the number of 2D correlated Cu spins, which provides a rough estimate
for $\xi_{2D}$. $\phi$ is the angle between $H$ and the octahedra tilt axis which is
normal to the DM plane (cf. inset Fig.~\ref{fig2}). In the PM phase $M_{DM}^{NO}$ equals
the full DM moment $M_{DM}$ and for $H\rightarrow \infty$ $M_{B}$ converges to $\pi/4
M_{DM}$. With $k$ used as a simple fit parameter Eq.~\ref{equ_mdmx} describes also the
remaining contribution $M_B$ of the non-AF-ordered fraction of $M_{DM}$ in the AF phase.
Our analysis has shown that $M_{DM}\simeq M_{DM}^{AF} + M_{DM}^{NO}$. A neglect of
$M_{B}$ would lead to inaccurate values for $M_{DM}^{AF}$, $H_{SF}$, and $J_\perp$, in
particular when $M_B\gtrsim M_{SF}$.

%\subsection{(Sr,Zn) doping dependence: $\bf H_{SF}$, $\bf M_{DM}$, and $\bf J_{\perp}$}

\begin{figure}[t]
\center{\includegraphics[width=0.95\columnwidth,angle=0,clip]{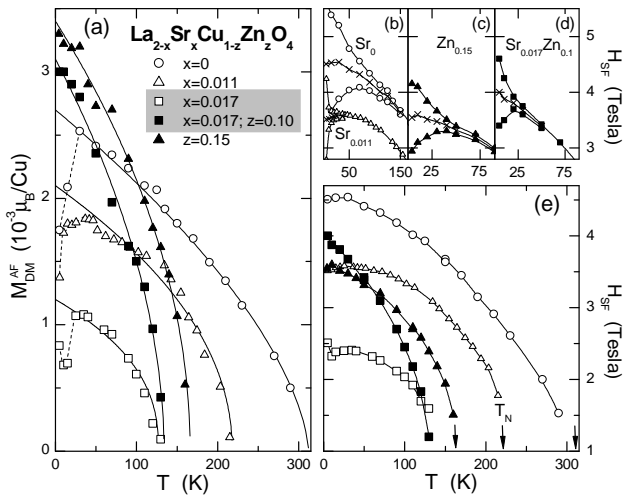}} \caption[]{Spin
flip parameters $M^{AF}_{DM}$ (a) and $H_{SF}$ (b-e) vs $T$ in \lsczo . (b-d) $H_{SF}$
for $dH/dt>0$ and $dH/dt<0$ as well as mean value ($\times$). (e) $H_{SF}$ mean
values.}\label{fig3}
\end{figure}

In Fig.~\ref{fig3} we show the $T$ dependence of $M^{AF}_{DM}$ and $H_{SF}$ of the five
samples studied (cf. Tab.~\ref{tab1}). In pure \lco\ $M^{AF}_{DM}$ increases monotonous
with decreasing $T$, and extrapolates to $M^{AF}_{DM}(0)=2.7 \times 10^{-3}$~$\rm
\mu_B/Cu$ for $T \rightarrow 0$~K. Pure Sr doping causes a drastic reduction of
$M^{AF}_{DM}$. In contrast, in \lczfo , though $T_N$ is strongly reduced, at low
temperatures $M^{AF}_{DM}$ (per Cu atom) becomes even larger than in \lco .
Interestingly, a similar behavior is observed for the Sr/Zn co-doped sample, where the
twofold role of Zn is to create spin vacancies and to reduce the mobility of the holes.

%
%
%Moreover, a further reduction of $M_{DM}$ is observed below $T\simeq 30$~K.
%In contrast to the solely Sr doped samples both Zn doped samples do not show
%the drop of $M_{DM}$ below $T\simeq 30$~K.

In \lco\ \HSF\ increases with decreasing $T$ and below $150$~K becomes hysteretic, with
its mean value saturating at 4.5~Tesla [Fig.~\ref{fig3}(b),(e)]. Pure Sr doping strongly
reduces \HSF\ as well as the hysteretic $T$ range. In particular for 1.7\% Sr only for
$T\leq 10$~K a clear field hysteresis is observed.
%with a remanent moment at 5~K which is $\sim$3 times larger than in \lco .
Both Zn doped samples show a relatively large field hysteresis [see (c),(d)]. Their
maximum mean value for \HSF\ is smaller than in \lco , but in view of their relatively
low $T_N$, \HSF\ is large if compared to \lssco .

As one can see in in Fig.~\ref{fig3}(a), in \lco\ and for pure Sr doping $M^{AF}_{DM}$
decreases below ~30~K. Though this looks similar for all $x$, there is a qualitative
difference. In \lco\ the hysteresis of the SF increases up to 2.2~Tesla at low $T$ and
$M(H)$ becomes strongly distorted which causes deviations from the fit and makes it
difficult to extract $M^{AF}_{DM}$. In contrast, in particular in the 1.7\% Sr doped
sample $M^{AF}_{DM}$ drops between 30~K and 15~K where $M(H)$ is reversible. Therefore,
we think that here the drop of $M^{AF}_{DM}$ is connected to the SFR. As $M^{AF}_{DM}
\propto$ AF order parameter, its drop in the SFR signals a degradation of the long range
order, consistent with the neutron diffraction data in Ref.~\onlinecite{Matsuda01a}.
Obviously, this effect is absent in both Zn doped samples. For the Sr/Zn doped sample
this evidences for a suppression of the SFR in \lsco\ by Zn.

\begin{figure}[t]
\center{\includegraphics[width=0.79\columnwidth,angle=0,clip]{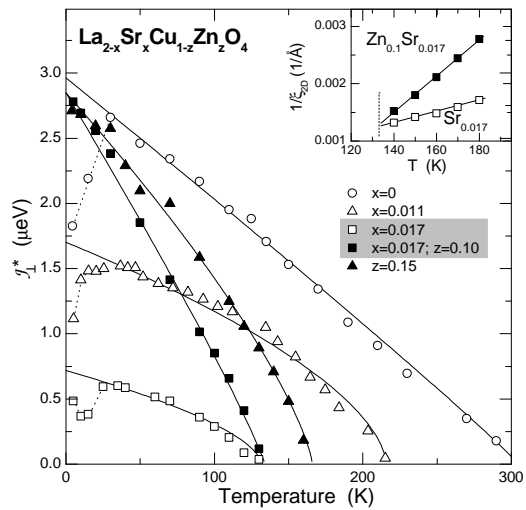}}
\caption[]{Interlayer coupling in \lsczo . Inset: Inverse correlation length vs $T$ for
both samples with $x=0.017$. Solid lines are guides to the eye.} \label{fig4}
\end{figure}

Following Eq.~\ref{equ_flip} we show in Fig.~\ref{fig4} the $T$-dependence of ${\mathcal
J^*_\perp} = M_{DM}^{AF} H_{SF}/S^2$ which we call the effective interlayer coupling,
where \HSF\ are the mean values in Fig.~\ref{fig3}(e). ${\mathcal J^*_\perp}$ accounts
for the effects of doping ($x,z$) and temperature which counter the interlayer
superexchange and only for $T \rightarrow 0$ ${\mathcal J^*_\perp}(0)=J_\perp$ (see
Tab.~\ref{tab1}). In \lco\ ${\mathcal J^*_\perp}$ increases with decreasing $T$ and for
$T\rightarrow 0$ we find ${\mathcal J^*_\perp}(0)=2.9$~$\mu$eV which is in good agreement
with 2.6~$\mu$eV for a single crystal with $T_N=240$~K in Ref.~\cite{Thio88}. As a
function of Sr doping ${\mathcal J^*_\perp}$ drastically decreases, and for $x=0.017$
${\mathcal J^*_\perp}(0)$ amounts to only 25\% of the value in \lco . If we compare this
to \lssczo\ which has exactly the same $T_N$, the huge difference is apparent. Here,
${\mathcal J^*_\perp}$ rises steeply and reaches almost the same value as in \lco .
\lczfo\ behaves similar.
With the hindsight that ${\mathcal J^*_\perp}(0)$ in \lco\ and the Zn doped samples is
about the same, it is evident from Eq.~\ref{equ_flip} that the larger DM moments in the
Zn doped samples have to result in correspondingly smaller critical fields \HSF\
[Fig.~\ref{fig3}(a),(e)]. To a certain extend the 15\% ($z=0.1$) and 25\% ($z=0.15$)
larger DM moments can be explained by an enhanced octahedra tilt angle
$\Phi$~\cite{tilt}. Local strain around the non-Jahn-Teller-active Zn sites might amplify
this effect.

%\subsection{Discussion of IC}
%
From Fig.~\ref{fig4} it is obvious that the mechanism for the suppression of the 3D AF
order is completely different for pure hole doping (Sr) on the one hand and non-magnetic
impurity doping (Zn) as well as mixed Sr/Zn doping on the other hand. In \lsco\ the hole
mobility increases rapidly with increasing $x$, and just a small concentration of mobile
holes reduces both, $T_N$ and ${\mathcal J^*_\perp}$. Co-doping with Zn reduces the hole
mobility~\cite{Huecker98c} and causes a drastic increase of ${\mathcal J^*_\perp}$, even
if $T_N$ is low. In particular for the two samples doped by 1.7\% Sr the influence of the
hole mobility is apparent. Our data clearly demonstrate that at fixed Sr content
$x\lesssim 0.02$ the interlayer coupling drastically increases if the hole mobility is
reduced.

Localized holes are supposed to suppress $T_N$ much stronger than Zn because a hole
located on an O-site ferromagnetically frustrates the AF Cu-O-Cu
exchange~\cite{Aharony88}. In contrast, a static spin vacancy does not perturb the AF
order of the surrounding Cu spins~\cite{Huecker02a}. If we compare now the 10\% Zn and
1.7\% Sr co-doped sample with a $\sim$18\% Zn doped sample, which would have the same
$T_N$ of $\sim$135~K, it is evident that 1.7\% localized holes suppress $T_N$ as
effective as 8\% Zn. This means that also for localized holes the effect of in-plane
ferromagnetic frustrations on $T_N$ is strong, while their effect on the interlayer
coupling is very weak (Fig.~\ref{fig4}), which is in contrast to the frustration
model~\cite{Aharony88}. Obviously the suppression of the interlayer coupling in \lsco\ is
closely connected to a high hole mobility (cf. Ref.~\onlinecite{Huecker98c} for $\rho$
values).

Lets focus on the correlation length of the two samples with 1.7\% Sr in the inset of
Fig.~\ref{fig4}. According to our analysis they have about the same $\xi_{2D}$ at $T_N$.
However, the sample with localized holes shows a much faster increases of $\xi_{2D}$ with
decreasing $T$ than the sample with mobile holes. In view of these results for $T \geq
T_N$ and the fact that the AF ordered moment at low $T$ stays much smaller in the purely
Sr doped sample, we think that mobile holes hinder $\xi_{2D}$ to develop and frustrate
the interlayer coupling in a way that cannot be explained with isolated localized holes.

Most of our observations presented in this paper can be naturally explained assuming
dynamic magnetic antiphase boundaries for $T\gtrsim 30$~K and static for $T\lesssim
30$~K. Evidence for static antiphase boundaries in the SFR was recently found by
Matsuda~et~al.~\cite{Matsuda01a}. Below 30~K an incommensurate AF order was detected
which coexists with the commensurate AF order. Dynamic antiphase domains were suggested
to explain the drastic reduction of the AF order parameter with increasing $x$~for
$30~K<T<T_N$~\cite{Suh98a}, but so far there is no direct evidence for that. Our
discussion does not depend on a particular domain form or on the formation of charge
stripes. We think that what is essential is the presence of mobile holes, as only these
holes can affect many Cu-O-Cu-bonds and are able to excite at least antiphase boundary
segments, i.e. strings of ferromagnetically frustrated Cu-O-Cu-bonds. In such an
environment a frustration of the interlayer coupling is inevitable, as two antiphased
domains in one $\rm CuO_2$-plane cannot AF couple simultaneously to the adjacent plane,
assuming that in particular in the case of dynamic antiphase boundaries 3D correlations
are absent.

The drastic reduction of $M^{AF}_{DM}$ in \lsco\ with increasing $x$ for $30{\rm
K}<T<T_N$ as well as its further reduction in the SFR can be explained in this framework
too. Just those regions of the $\rm CuO_2$ planes with AF interlayer coupling contribute
to the spin flip ($M_{SF}$), while regions with ferromagnetically frustrated interlayer
coupling can be aligned continuously with increasing field ($M_{B}$). Both contributions
must result from correlated regions with relatively large $\xi_{2D}$ as it is possible to
align the full DM moment at any temperature (cf. Fig.~1 for high $H$ up to $H=14$~Tesla).
Interestingly, no significant change of the mean value $H_{SF}$ is observed upon the
transition into the SFR, whereas the decrease in $M_{SF}$ and a corresponding increase in
$M_{B}$ was very clear. Obviously, just the fraction of regions with AF interlayer
coupling decreases and not their interlayer coupling. This suggests that in the SFR
coupled and frustrated regions are spatially separated, which is consistent with the
observation of two $\mu SR$ frequencies in the SFR of the \lssco\ sample~\cite{Wagener98}
and recent neutron diffraction data in Ref.~\onlinecite{Matsuda01a}. The reduction of
$M^{AF}_{DM}$ in the SFR could be the consequence of a more destructive or a more even
distribution of antiphase boundaries when they become statically ordered. We think that
on either side of an in-plane antiphase boundary the DM moments point into opposite
directions. At first glance this looks like DM moments cancel out, but it is important to
consider both, their arrangement in the planes as well as between the planes. In-plane
phase boundaries cause jumps in the AF in-plane phase but do not necessarily destabilize
in-plane correlations, whereas they certainly result in a frustration of the interlayer
coupling.

On the basis of our experimental data we come to the conclusion that the drastic
suppression of the interlayer coupling and hence of $T_N$ in \lsco\ is a real {\it
magnetic decoupling} of the $\rm CuO_2$ planes due to the ferromagnetic frustration of
the interlayer coupling by in-plane magnetic antiphase boundaries. We have shown that an
important condition for this effect is a high hole mobility, and that the ferromagnetic
frustration effect of localized holes on the interlayer coupling is small.

%\begin{acknowledgments}
%
We gratefully acknowledge helpful discussions with M.~Vojta and F.~Essler. Work at
Brookhaven was supported by the Material Science Division, Department of Energy, under
Contract No. DE-AC02-98CH10886.


\begin{thebibliography}{14}
\expandafter\ifx\csname natexlab\endcsname\relax\def\natexlab#1{#1}\fi
\expandafter\ifx\csname bibnamefont\endcsname\relax
  \def\bibnamefont#1{#1}\fi
\expandafter\ifx\csname bibfnamefont\endcsname\relax
  \def\bibfnamefont#1{#1}\fi
\expandafter\ifx\csname citenamefont\endcsname\relax
  \def\citenamefont#1{#1}\fi
\expandafter\ifx\csname url\endcsname\relax
  \def\url#1{\texttt{#1}}\fi
\expandafter\ifx\csname urlprefix\endcsname\relax\def\urlprefix{URL }\fi
\providecommand{\bibinfo}[2]{#2} \providecommand{\eprint}[2][]{\url{#2}}

\bibitem[{\citenamefont{Niedermayer et~al.}(1998)\citenamefont{Niedermayer}}]{Niedermayer98}
\bibinfo{author}{\bibfnamefont{C.}~\bibnamefont{Niedermayer}} et al.,
\bibinfo{journal}{Phys.\ Rev.\ Lett.}
\textbf{\bibinfo{volume}{80}}, \bibinfo{pages}{3843} (\bibinfo{year}{1998}).

\bibitem[{\citenamefont{Aharony et~al.}(1988)\citenamefont{Aharony}}]{Aharony88}
\bibinfo{author}{\bibfnamefont{A.}~\bibnamefont{Aharony}} et al.,
\bibinfo{journal}{Phys.\ Rev.\ Lett.}
\textbf{\bibinfo{volume}{60}}, \bibinfo{pages}{1330} (\bibinfo{year}{1988}).

\bibitem[{\citenamefont{Thio et~al.}(1988)\citenamefont{Thio}}]{Thio88}
\bibinfo{author}{\bibfnamefont{T.}~\bibnamefont{Thio}} et al.,
\bibinfo{journal}{Phys.\ Rev.} \textbf{\bibinfo{volume}{B~38}},
\bibinfo{pages}{905} (\bibinfo{year}{1988}).

\bibitem[{\citenamefont{Chou et~al.}(1993)\citenamefont{Chou}}]{Chou93}
\bibinfo{author}{\bibfnamefont{F.~C.} \bibnamefont{Chou}} et al.,
\bibinfo{journal}{Phys.\ Rev.\ Lett.} \textbf{\bibinfo{volume}{71}},
\bibinfo{pages}{2323} (\bibinfo{year}{1993}).

\bibitem[{\citenamefont{Borsa et~al.}(1995)\citenamefont{Borsa}}]{Borsa95}
\bibinfo{author}{\bibfnamefont{F.}~\bibnamefont{Borsa}} et al.,
\bibinfo{journal}{Phys.\ Rev.} \textbf{\bibinfo{volume}{B~52}},
\bibinfo{pages}{7334} (\bibinfo{year}{1995}).

\bibitem[{\citenamefont{H\"ucker et~al.}(1998)\citenamefont{H\"ucker}}]{Huecker98c}
\bibinfo{author}{\bibfnamefont{M.}~\bibnamefont{H\"ucker}} et al.,
\bibinfo{journal}{Phys.\ Rev.}
\textbf{\bibinfo{volume}{B~59}}, \bibinfo{pages}{R725} (\bibinfo{year}{1999}).

\bibitem[{\citenamefont{Matsuda et~al.}(2002)\citenamefont{Matsuda}}]{Matsuda01a}
\bibinfo{author}{\bibfnamefont{M.}~\bibnamefont{Matsuda}} et al.,
\bibinfo{journal}{Phys.\ Rev.}
\textbf{\bibinfo{volume}{B~65}}, \bibinfo{pages}{134515} (\bibinfo{year}{2002}).

\bibitem[{\citenamefont{Cho et~al.}(1993)\citenamefont{Cho}}]{Cho93}
\bibinfo{author}{\bibfnamefont{J.~H.} \bibnamefont{Cho}} et al.,
\bibinfo{journal}{Phys.\ Rev.\ Lett.} \textbf{\bibinfo{volume}{70}},
\bibinfo{pages}{222} (\bibinfo{year}{1993}).

\bibitem[{\citenamefont{Korenblit et~al.}(1999)\citenamefont{Korenblit}}]{Korenblit99}
\bibinfo{author}{\bibfnamefont{I.~Y.} \bibnamefont{Korenblit}} et al.,
\bibinfo{journal}{Phys.\ Rev.} \textbf{\bibinfo{volume}{B~60}},
\bibinfo{pages}{R15017} (\bibinfo{year}{1999}).

\bibitem[{\citenamefont{Suh et~al.}(1998)\citenamefont{Suh}}]{Suh98a}
\bibinfo{author}{\bibfnamefont{B.~J.} \bibnamefont{Suh}} et al.,
\bibinfo{journal}{Phys.\ Rev.\ Lett.} \textbf{\bibinfo{volume}{81}},
\bibinfo{pages}{2791} (\bibinfo{year}{1998}).

\bibitem[{\citenamefont{Neto}(1998)}]{Neto98b}
\bibinfo{author}{\bibfnamefont{A.~H.~C.} \bibnamefont{Neto}} et al.,
%  \bibinfo{author}{\bibfnamefont{A.~V.} \bibnamefont{Balatsky}},
  \bibinfo{journal}{cond-mat} \textbf{\bibinfo{volume}{/9805273}}
  (\bibinfo{year}{1998}).

\bibitem[{\citenamefont{Thio and Aharony}(1994)}]{Thio94}
\bibinfo{author}{\bibfnamefont{T.}~\bibnamefont{Thio}} \bibnamefont{and}
  \bibinfo{author}{\bibfnamefont{A.}~\bibnamefont{Aharony}},
  \bibinfo{journal}{Phys.\ Rev.\ Lett.} \textbf{\bibinfo{volume}{73}},
  \bibinfo{pages}{894} (\bibinfo{year}{1994}).

\bibitem{tilt} $T_{HT}$ of the strucutral transition HTT$\rightarrow$LTO
changes by $8 {\rm K}/0.01$Zn and by $-25 {\rm K}/0.01$Sr. Hence, $M_{DM} \propto \Phi
\propto [(T_{HT}-T)^{0.62}]^{1/2}$ increases by $\sim$3\% ($z=0.10$) and $\sim$7\%
($z=0.15$) at $T=0$~K with respect to \lco ~\cite{Thio88,Buechner}.

\bibitem[{\citenamefont{H\"ucker and B\"uchner}(2002)\citenamefont{H\"ucker}}]{Huecker02a}
\bibinfo{author}{\bibfnamefont{M.}~\bibnamefont{H\"ucker}} \bibnamefont{and}
\bibinfo{author}{\bibfnamefont{B.}~\bibnamefont{B\"uchner}},
\bibinfo{journal}{Phys.\ Rev.}
\textbf{\bibinfo{volume}{B~65}}, \bibinfo{pages}{214408} (\bibinfo{year}{2002}), and
references therein.

\bibitem{Wagener98} W.~Wagener, Ph.D. thesis, TU Braunschweig, 1998

\end{thebibliography}
\end{document}